\documentclass[pra,aps,superscriptaddress,twocolumn,color]{revtex4-2}

\usepackage{amsmath}
\usepackage{amsfonts}
\usepackage{bm}
\usepackage{graphicx}
\usepackage{array,color}
\usepackage{braket}
\usepackage[table]{xcolor}
\usepackage{xcolor}
\usepackage{chemformula}
\usepackage{hyperref}
\usepackage{cleveref}

\begin{document}
	\title{Emergent aperiodicity in Bose-Bose mixtures induced by spin-dependent periodic potentials}
	
	\author{Abid Ali}
	\affiliation{First Affiliated Hospital of Xi'an Jiaotong University, Xi'an 710049, People’s Republic of China}
	\affiliation{MOE Key Laboratory for Nonequilibrium Synthesis and Modulation of Condensed Matter, and Shaanxi Key Laboratory of Quantum Information and Quantum Optoelectronic Devices, School of Physics, Xi’an Jiaotong University, Xi’an 710049, People’s Republic of China}
	
	\author{Pei Zhang}
	\affiliation{MOE Key Laboratory for Nonequilibrium Synthesis and Modulation of Condensed Matter, and Shaanxi Key Laboratory of Quantum Information and Quantum Optoelectronic Devices, School of Physics, Xi’an Jiaotong University, Xi’an 710049, People’s Republic of China}
	
	\author{Hiroki Saito}
	\affiliation{ Department of Engineering Science, University of Electro-Communications, Tokyo 182-8585, Japan}
	
	\author{Yong-Chang Zhang}
	\email{zhangyc@xjtu.edu.cn}
	\affiliation{MOE Key Laboratory for Nonequilibrium Synthesis and Modulation of Condensed Matter, and Shaanxi Key Laboratory of Quantum Information and Quantum Optoelectronic Devices, School of Physics, Xi’an Jiaotong University, Xi’an 710049, People’s Republic of China}

	\date{\today}
	
	\begin{abstract}
		
We study the ground-state and low-lying metastable phases of repulsive binary Bose-Einstein condensates confined in twisted, spin-dependent periodic optical lattices. For balanced mixtures, weak intercomponent interactions yield a fourfold momentum-space symmetry dictated by the lattice geometry. Increasing the coupling strength leads to the emergence of additional momentum peaks that combine with the lattice-induced structure to produce an eightfold rotationally symmetric pattern, signaling quasicrystalline order. At intermediate interactions, global phase separation suppresses this quasicrystalline state; however, at stronger coupling, local phase separation gives rise to a long-lived metastable phase in which the eightfold symmetry is restored. In this regime, a secondary ring of dominant momentum peaks appears at smaller wave vectors, indicating longer-wavelength density modulations and a crossover from lattice-dominated to interaction-driven quasicrystalline order. In contrast, imbalanced mixtures form partially miscible density clusters with eightfold-symmetric aperiodic patterns only at intermediate coupling, while stronger interactions drive global phase separation and permanently destroy quasicrystalline order. Real-time simulations demonstrate that these aperiodic structures are dynamically stable and experimentally accessible. Our results show that quasicrystalline order can emerge in binary condensates without explicitly aperiodic lattices and reveal population balance as a key ingredient for stabilizing quantum quasicrystals.

\end{abstract}
\maketitle
\section{Introduction}
	
Unlike classical crystals, which exhibit both rotational and translational symmetries, quasicrystals (QCs) defy conventional symmetry rules and lack periodicity. QCs often feature rotational symmetries—such as fivefold, sevenfold, eightfold, ninefold, tenfold, and twelvefold—that are forbidden in periodic crystals~\cite{1,2,Steurer}. Positioned at the boundary between order and disorder, QCs represent a fascinating phase of matter that challenges traditional crystallography, offering new insights into structural complexity and symmetry. These aperiodic phases have attracted significant attention across various fields, uncovering new avenues for exploring exotic states of matter~\cite{Gomila2014PRE,6,Deng2019PRL,Heinonen2019PRA,Cinti2022PRB,Cinti2023PRL} and their potential applications in a wide range of areas, from materials science~\cite{Inoue2008,Edagawa,Simonov2020review} to quantum optics~\cite{Freedman2006,Gao2012,Agrawal2013}, toplogical physics~\cite{Kraus2012PRL,Wang2019,Else2021prx,Weidemann2022Nature}, quantum transport~\cite{Chawla2019,Xing2022}, and quantum computation~\cite{Irwin2022}.
	
In recent years, the study of QCs in ultracold atomic systems has gained significant momentum, offering valuable insights into their formation and stabilization under controlled conditions. Leveraging the high controllability of ultracold atoms, various theoretical approaches have been proposed to explore QCs in these systems~\cite{Mace2016Crystals,Gautier2021PRL,Khazali2021prr}. For example, in dipolar quantum gases, the interplay between spin-orbit coupling and long-range dipole-dipole interactions plays a crucial role in stabilizing complex quantum crystalline and quasicrystalline ground states~\cite{7}. Path-integral Monte Carlo simulations have demonstrated that moderate quantum fluctuations can preserve quasicrystalline order, enabling small yet finite superfluidity even in open crystalline lattices~\cite{8}. 
Furthermore, due to their remarkable quantum properties compared to classical materials, degenerate atomic gases, such as Bose-Einstein condensates (BECs), provide an ideal platform for investigating exotic quantum phases like supersolids and high-order QCs. For instance, a superfluid-quasicrystal state with minimal fivefold rotational symmetry has been predicted in BECs confined in a tilted superlattice potential~\cite{9}. The transition from a homogeneous BEC to a quasicrystalline phase, driven by either external trapping potentials with typical QCs symmetries~\cite{Gautier2021PRL} or by collective superradiant light scattering in a cavity-QED setup~\cite{Farokh2019PRL}, has been observed. More recently, the exploration of quantum states of dipolar BECs in the presence of elaborated QC-shape confinement further highlights the potential of these systems for probing quantum phase transitions in QCs and other aperiodic states of matter~\cite{Cinti2024PRL}.
	
The first experimental realization of QCs in ultracold atoms was achieved using eightfold rotationally symmetric optical lattices~\cite{11}. In a subsequent study, the localization transition in the ground state induced by such an eightfold symmetric quasicrystalline optical lattice was explored~\cite{localization2020}. This work provided a groundbreaking platform for investigating the properties of QCs and offered valuable insights into the nature of QCs as well as the behavior of light-atom interacting systems. However, unlike natural QCs, the quasicrystallization observed in these experiments is not dynamically emergent or spontaneous. Instead, it is driven by the application of eightfold rotationally symmetric optical lattice potentials to the atoms through external lasers.
	
It remains desirable to generate QCs through a self-organization process, rather than simply imprinting the aperiodic symmetry of the external confinement onto the BECs~\cite{Cinti2023PRL}. To explore this, we consider a binary BEC system subjected to two-dimensional component-dependent periodic potentials and investigate its ground-state properties. In contrast to the previously mentioned configuration of a single-component BEC subjected to an eightfold rotationally symmetric potential, where spatial translational symmetry is broken solely by the external aperiodic confinement, each species in our system experiences a spin-dependent periodic potential. In this scenario, the trapping potential actually works to preserve, rather than break, the translational symmetry. These potentials, generated using laser beams with selected wave numbers  \( k \) and polarizations~\cite{spin1,spin2,spin3,spin4,spin5,spin6,spin7,twisted}, introduce complexity through the interplay between intra-component interactions \( g_{11}, g_{22} \) and intercomponent interactions \( g_{12} \). The competition between interaction scales, which is known to drive quasicrystallization in soft matter systems~\cite{12,13,14,15,16,17,18}, is expected to similarly influence the formation of QCs in atomic systems.
	
By analyzing the ground-state behavior of this binary BEC system, we find that atom-atom contact interactions can lead to remarkable quasicrystalline density profiles within the periodic background potentials. For a balanced mixture (i.e., when the two components have the same particle number), aperiodic quasicrystalline density modulations emerge, accompanied by an immiscible-miscible phase transition. In the case of imbalanced mixtures, we show that the two components are partially overlap even though $g_{12}>g$; this occurs due to density modulation within each component, which alters the global mixing energy. This partially miscible phase self-organizes into aperiodically modulated density clusters exhibiting eight-fold rotational symmetry.
	
The rest of this paper is organized as follows. In Sec.~II, we describe the system using coupled Gross-Pitaevskii equations with a spin-dependent periodic potential. In Sec.~III, we present numerical results for both balanced and imbalanced mixtures, focusing on the ground-state density profile. In Sec.~IV, we exhibit real-time dynamics of the binary system. Finally, Sec.~V summarizes the key findings.
	
\section{Theoretical Model of the binary BECs}
	
The model considered here describes a two-dimensional binary Bose-Einstein condensate at zero temperature, subjected to spin-dependent two-dimensional periodic potentials. Within the mean-field approximation, the system is characterized by the macroscopic wave functions $\Psi_1$ and $\Psi_2$, whose dynamics are governed by the coupled Gross-Pitaevskii (GP) equations
\begin{subequations}\label{1}
		\begin{align}
			i \hbar \frac{\partial \Psi_1}{\partial t} &= \Big(-\frac{\hbar^2 \nabla^2}{2 m_1} + V_1 + V_{\rm trap} + g_{11} |\Psi_1|^2 + g_{12} |\Psi_2|^2 \Big) \Psi_1, \label{GP_a} \\
			i \hbar \frac{\partial \Psi_2}{\partial t} &= \Big(-\frac{\hbar^2 \nabla^2}{2 m_2} + V_2 + V_{\rm trap} + g_{22} |\Psi_2|^2 + g_{21} |\Psi_1|^2 \Big) \Psi_2. \label{GP_b}
		\end{align}
\end{subequations}
For simplicity, we set $m_1 = m_2 = 1$ and $\hbar = 1$, with $m_1$ and $m_2$ denoting the atomic masses of components 1 and 2, respectively. The wave functions are normalized such that $\int d\mathbf{r} \, |\Psi_j(\mathbf{r}, t)|^2 = N_j$, where $N_j$ is the number of atoms in component $j = 1,2$. The interaction coefficients are defined as $g_{jj'} = 4 \pi \hbar^2 a_{jj'} / m$, where $a_{jj'}$ is the $s$-wave scattering length between components $j$ and $j'$.
	
The external potential acting on the condensates consists of a spin-dependent periodic lattice and a circular confinement trap. The periodic potentials are given by
	\begin{subequations}
		\begin{align}
			V_1(x,y) &= U_1 \big(\cos^2(k x) + \cos^2(k y)\big), \\
			V_2(x',y') &= U_2 \big(\cos^2(k x') + \cos^2(k y')\big),
		\end{align}
	\end{subequations}
where $(x',y')$ denotes a coordinate system rotated by an angle $\theta$ relative to $(x,y)$:
	\begin{equation}\label{4}
		\begin{pmatrix} x' \\ y' \end{pmatrix} =
		\begin{pmatrix} \cos \theta & -\sin \theta \\ \sin \theta & \cos \theta \end{pmatrix}
		\begin{pmatrix} x \\ y \end{pmatrix}.
	\end{equation}
Here, $U_1 = -U_2 \equiv U$ defines the lattice depth for the two components.
	
To confine the condensates within a finite region, we introduce a soft circular trap,
	\begin{equation}\label{Vtrap}
		V_{\rm trap}(r) = \frac{V_0}{2} \Big(1 + \tanh\frac{r - R_t}{w_t}\Big),
	\end{equation}
where $r = \sqrt{x^2 + y^2}$, $V_0$ is the trap amplitude, $R_t$ is the trap radius, and $w_t$ characterizes the smoothness of the trap edge. The circular trap serves two important purposes in the context of the twisted bilayer potential: (i) it confines the condensate within the finite numerical domain, preventing the wave function from interacting with the simulation boundaries and thereby avoiding edge effects, and (ii) it preserves the rotational symmetry of the system, ensuring that the rotated lattice does not induce distortions in the density peaks in momentum space due to asymmetric boundary interactions. This form provides a nearly flat potential in the central region with smoothly varying edges, allowing the atomic density to remain uniform at the center while minimizing reflections and other numerical artifacts at the boundaries.

\begin{figure}[tb]
	\includegraphics[width=7.9cm]{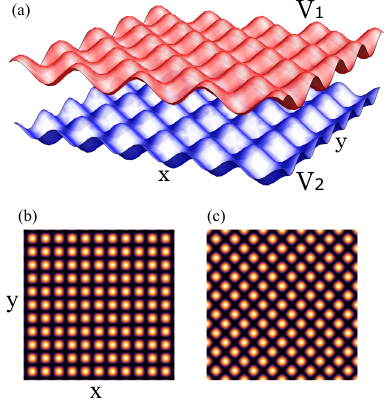}
	\caption{
		(a) Schematic illustration of two independent periodic optical lattice potentials, $V_1$ for component 1 and $V_2$ for component 2. 
		Panels (b) and (c) show the corresponding spatial profiles of $V_1$ and $V_2$ along the $x$ and $y$ directions, highlighting the relative orientation of the square lattices. This configuration enables independent manipulation of atoms in different spin components via spin-dependent optical lattices.
		\label{fig:fig1}}
\end{figure}

The spin-dependent periodic potentials $V_1(x, y)$ and $V_2(x', y')$ for component 1 and 2, respectively, with a fixed rotation angle of $\theta = \pi / 4$, are applied to the two-component BEC system as shown in Fig.~\ref{fig:fig1}(a). Recent experimental realization of such component-dependent periodic potentials in binary BEC systems has been demonstrated in \cite{twisted}, where the wavelengths of the lattice beams are precisely adjusted to the tune-out wavelengths. Consequently, atoms in state $|1\rangle$ are only diffracted by $V_1(x, y)$, while atoms in state $|2\rangle$ experience only $V_2(x', y')$.

It is worth highlighting that this twisted bilayer square lattice setting differs from the eightfold rotationally symmetric optical potentials used in Refs.~\cite{Gautier2021PRL,Cinti2024PRL,11}, even though the latter are similarly generated by the superposition of two square lattices with a relative rotational angle of $\pi/4$. In the configurations considered in Refs.~\cite{Gautier2021PRL,Cinti2024PRL,11}, a single-component BEC is subjected to overall aperiodic confinements with eightfold rotational symmetry. However, the two-layer potentials $V_{1,2}$ discussed in this work are spin-dependent and applied to two different species, with each species experiencing only one periodic square optical lattice. In the following, we will demonstrate that the interplay between these periodic external confinements and the atom-atom contact interactions can lead to symmetry breaking, resulting in emergent aperiodicity in the BEC density profiles. This behavior contrasts sharply with the quasicrystallization observed in Refs.~\cite{Gautier2021PRL,Cinti2024PRL,11}, which is solely induced by the external aperiodic potentials.
	
\section{Ground state properties}

\begin{figure*}[tb]
	\includegraphics[scale=1.29]{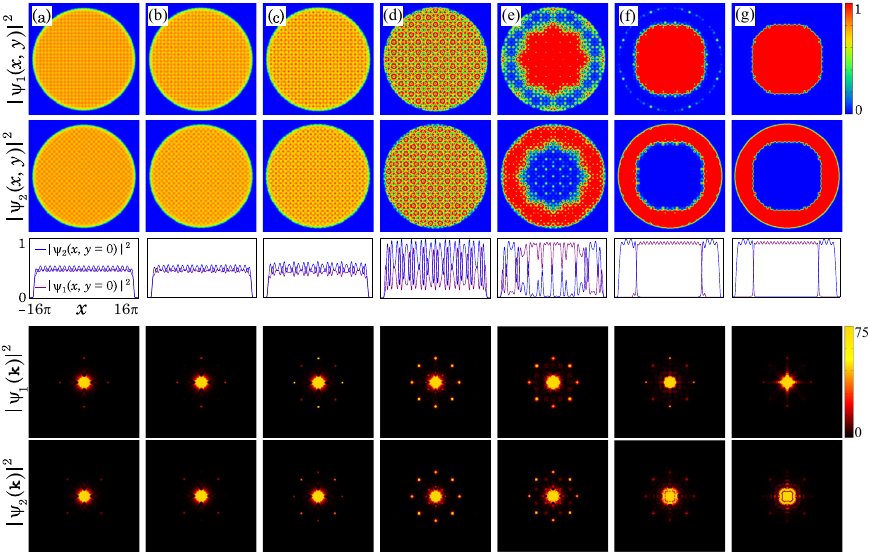}
	\caption{
		Ground-state density distributions of components~1 and~2 in a balanced binary BEC. The intra-component interactions are fixed at $g_{11} = g_{22} \equiv g = 15$, with lattice depth $U = 1.2$. The circular box trap parameters are $V_0 = 50$, $R_t = 15\pi$, and $w_t = 2$. The middle panels show line profiles along $y = 0$, $|\Psi_j(x, y=0)|^2$, while the bottom panels display the corresponding momentum-space distributions. Panels (a)–(f) correspond to intercomponent interaction strengths $g_{12} = 1.5$, $6$, $10.5$, $15.15$, $15.25$, $15.35$, and $15.65$, respectively. Each real-space panel spans a spatial domain of $(32\pi)^2$, and each momentum-space panel spans $(3\pi)^2$.}
	\label{fig:fig3}
\end{figure*}
\begin{figure*}[tb]
	\includegraphics[scale=0.7]{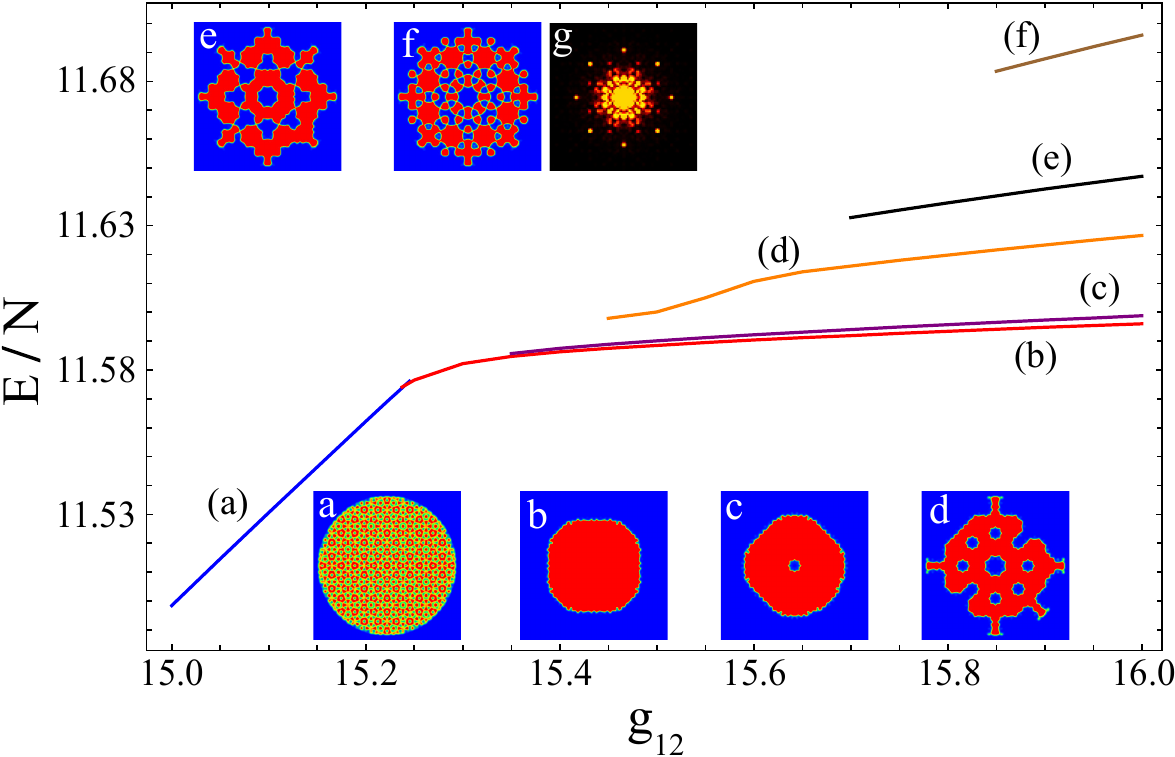}
	\caption{Energies of the ground and metastable states as a function of the intercomponent interaction strength $g_{12}$ for the balanced mixture. Solid lines indicate the energies of the respective states. Insets show the real-space density distribution of component~1 for each case. Other parameters are the same as in Fig.~\ref{fig:fig3}. Real-space densities are plotted over a spatial domain of $(32\pi)^2$, and the corresponding momentum-space distribution for the case shown in panel~(f) is displayed in panel~(g), spanning $(2\pi)^2$.}
	\label{fig:meta_stable}
\end{figure*}

\begin{figure}[tb]
	\includegraphics[scale=1.22]{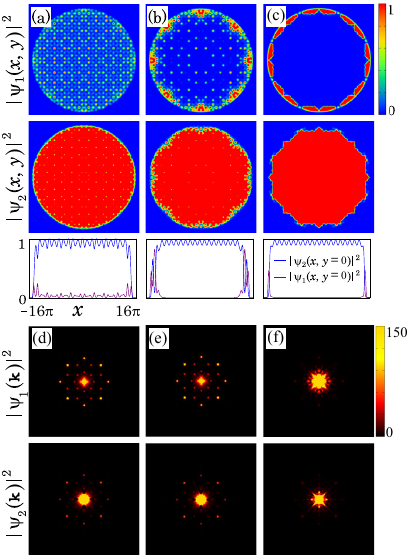}
	\caption{Ground-state density distributions of an imbalanced binary BEC in a spin-dependent periodic lattice. Panels (a)–(c) show the system for increasing intercomponent interactions $g_{12} = 15.15$, $15.3$, and $15.9$, respectively. The bottom panels display line profiles along $y=0$, $|\Psi_j(x, y=0)|^2$, and the corresponding momentum-space distributions. For moderate $g_{12}$ in panels (a) and (b), the condensates form partially miscible quasicrystalline density clusters with eightfold rotational symmetry, whereas for stronger $g_{12}$ in panel (c), the components undergo global phase separation and the quasicrystalline signature disappears. Other parameters are the same as in Fig.~\ref{fig:fig3}. ach real-space panel spans a spatial domain of $(32\pi)^2$, and each momentum-space panel spans $(3\pi)^2$.}
	\label{fig:fig2}
\end{figure}

We consider a two-dimensional binary BECs subjected to spin-dependent periodic potentials and a soft circular trapping potential, as described in Eq.~(\ref{Vtrap}). The imaginary-time propagation of Eq.~(\ref{1}) is employed to determine the ground state, where the imaginary unit \(i\) is replaced with \(-1\). Numerically, the system is discretized on a uniform Cartesian grid with spatial resolution \(dx = dy = 2\pi/64\) and a time step \(dt = 0.001\). The circular trap ensures smooth confinement within the simulation box, minimizing boundary artifacts while maintaining a nearly uniform density in the central region. To break the symmetry of the system and avoid metastable configurations, a small random perturbation is added at each grid point of the initial state. Throughout the simulations, we assume equal intra-component interactions, $g_{11} = g_{22} \equiv g$, and a total atom number $N = N_1 + N_2$.  

\begin{figure}[tb]
	\includegraphics[scale=0.5]{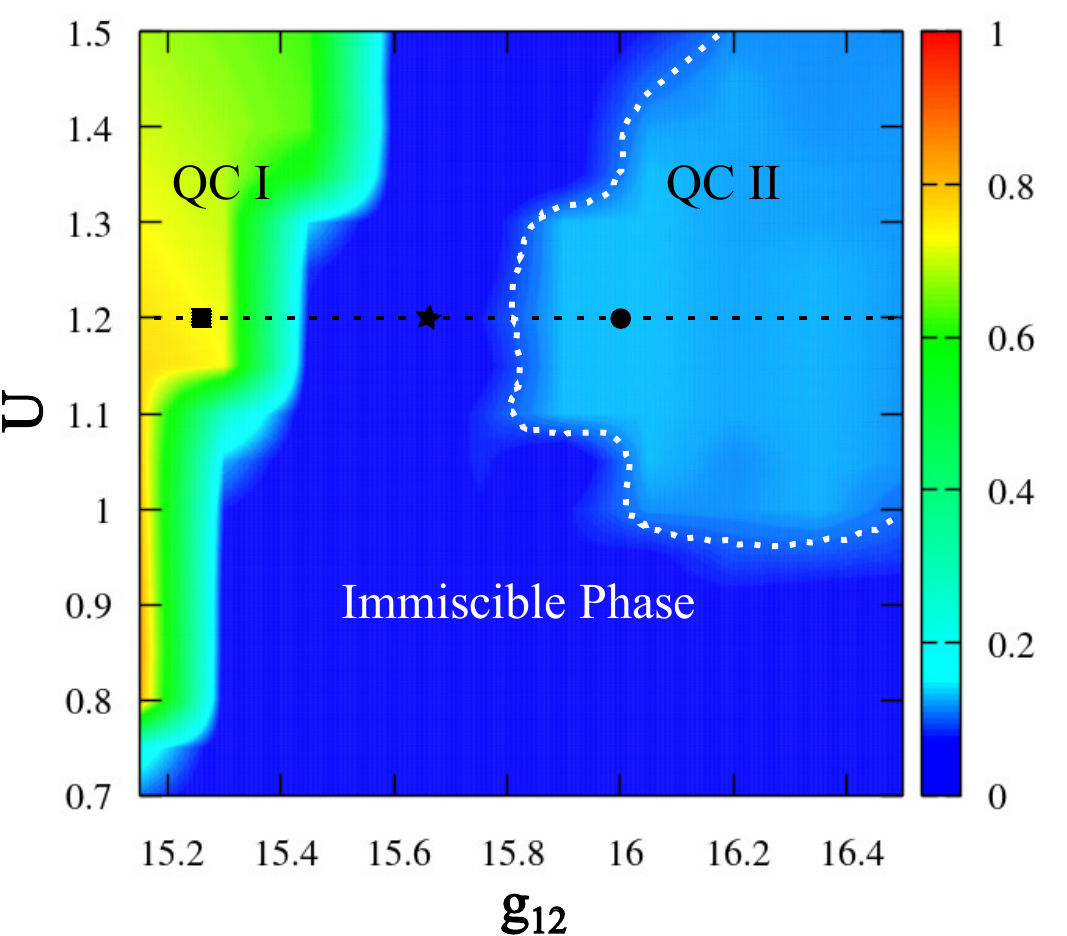}
	\caption{Phase diagram of the balanced binary BEC in the $(g_{12}, U)$ plane, obtained from the overlap parameter $\mathcal{O}$ defined in Eq.~\eqref{eq:overlap}. The black square marks the parameters of Fig.~\ref{fig:fig3}(e), corresponding to a quasicrystalline phase (QC I) where the system begins to segregate into large spatial domains while retaining eightfold symmetry in momentum space. The black star indicates the globally immiscible phase shown in Fig.~\ref{fig:fig3}(f), where $\mathcal{O}=0$ and quasicrystalline order is lost. The black circle denotes the metastable quasicrystalline regime (QC II) of Fig.~\ref{fig:meta_stable}(f), where density modulations re-emerge and the momentum distribution recovers eightfold rotational symmetry despite local phase separation.}
	\label{fig:phase-balance}
\end{figure}

\begin{figure}[tb]
		\includegraphics[scale=0.425]{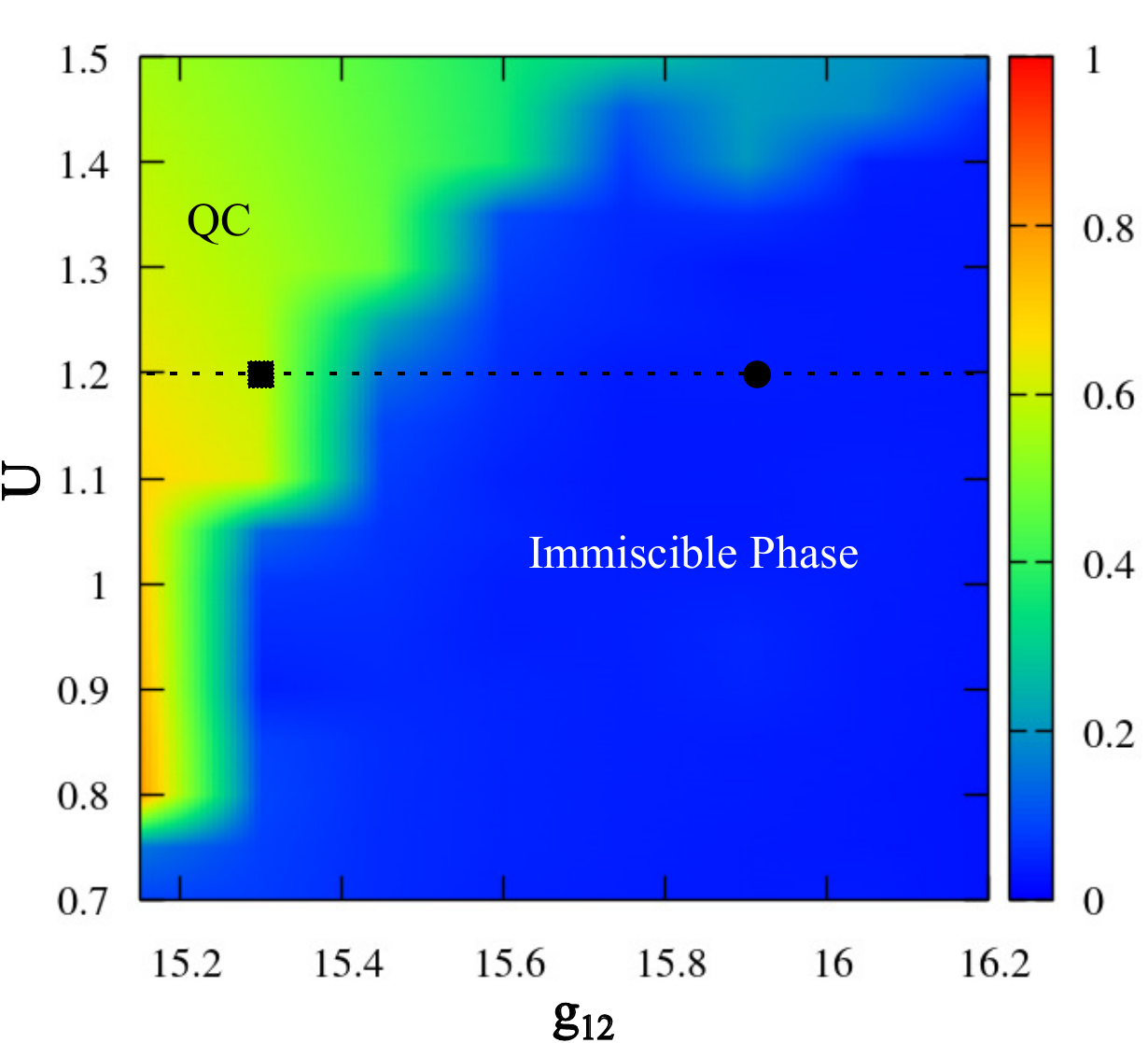}
\caption{Phase diagram of the imbalanced binary BEC in the $(g_{12}, U)$ plane, obtained from the overlap parameter $\mathcal{O}$. The black square marks the coexistence phase exhibiting quasicrystalline order (Fig.~\ref{fig:fig2}(b)), while the black circle indicates the fully immiscible phase (Fig.~\ref{fig:fig2}(c)).}
\label{fig:phase-imbalance}
\end{figure}

After setting up the system as described, we first examine the case of a balanced mixture, where the populations of components~1 and~2 are equal. Figure~\ref{fig:fig3} shows the corresponding ground-state density profiles $|\Psi_1|^2$ and $|\Psi_2|^2$, which exhibit aperiodic patterns arising from the combined effects of the intercomponent interaction $g_{12}$ and the spin-dependent periodic optical lattice potentials $V_1$ and $V_2$. The external potentials act independently on each component, while the interaction strength $g_{12}$ determines the degree of spatial overlap between them.
	
In Fig.~\ref{fig:fig3}(a), the real-space density distributions follow the underlying square-lattice geometry, with component~2 rotated according to the potential $V_2(x',y')$. As the intercomponent interaction $g_{12}$ increases, the coupling between the two layers strengthens and progressively distorts the original lattice order. For sufficiently strong coupling (e.g., $g_{12}=10.5$), the periodic density modulations are replaced by aperiodically structured patterns in both components, as illustrated in Fig.~\ref{fig:fig3}(c). The middle panels of Fig.~\ref{fig:fig3} show the density profiles along $y=0$, where both components exhibit irregular modulations that become increasingly pronounced with growing $g_{12}$.

At weak coupling ($g_{12}=1.5$), the momentum-space distributions exhibit four low-intensity peaks per component, consistent with the $\pi/4$ rotation between the two square lattices. These lattice-related peaks set the dominant modulation length scale imposed by the optical potentials. As $g_{12}$ increases, the four primary peaks become more pronounced, and at $g_{12}=10.5$ (Fig.~\ref{fig:fig3}(c)) weak secondary peaks appear, diagonally in component~1 and off-diagonally in component~2; these secondary peaks signal the onset of quasicrystalline correlations, although their intensity is still small. With a further increase to $g_{12}=15.15$ (Fig.~\ref{fig:fig3}(d)), the secondary peaks strengthen and, together with the four lattice-induced peaks, produce eight distinct peaks that define a clear eightfold rotational symmetry.

With further increasing intercomponent interaction strength, the balance between mixing and segregation changes qualitatively. At $g_{12}=15.25$ (Fig.~\ref{fig:fig3}(e)), the two condensates start to separate into extended domains; nevertheless, the momentum distributions of both components still display a clear eightfold rotational symmetry. Upon slightly increasing the coupling to $g_{12}=15.35$ (Fig.~\ref{fig:fig3}(f)), global phase separation becomes dominant, accompanied by a noticeable reduction in the intensity of the secondary momentum peaks. At even stronger coupling, $g_{12}=15.65$ (Fig.~\ref{fig:fig3}(g)), the system forms fully separated macroscopic domains, and the secondary peaks vanish entirely, leaving no signature of quasicrystalline order in the ground state. This evolution from fourfold lattice-induced symmetry at weak coupling to eightfold rotational symmetry at intermediate coupling provides direct evidence of quasicrystalline order in a binary BEC realized here without the need for an externally imposed quasiperiodic lattice~\cite{Gautier2021PRL,Cinti2024PRL,11}.

In contrast, Fig.~\ref{fig:meta_stable} presents the energies of the ground and metastable states as a function of the intercomponent interaction strength $g_{12}$. Panels (a) and (b) correspond to the ground state, whereas panels (c)--(f) show distinct metastable configurations that evolve differently as $g_{12}$ is increased. These metastable states are characterized by locally phase-separated regions with alternating, finely modulated density patterns.

Notably, the momentum-space distributions in panels (b) and (c) exhibit an abrupt disappearance of eightfold symmetry as a consequence of global phase separation. In contrast, for the locally phase-separated configurations shown in panels (d)--(f), the alternating, finely modulated density regions lead to a re-emergence of eightfold symmetry in momentum space, accompanied by sharper and more intense secondary peaks. Panel (g) displays the momentum-space distribution corresponding to the real-space density in panel (f), demonstrating that quasicrystalline order can persist robustly in metastable states. Nevertheless, these configurations do not correspond to the true ground state of the system.

The re-emergence of eightfold symmetry in the metastable configuration (Fig.~\ref{fig:meta_stable}(f)) is driven by a change in the characteristic length scale of density modulation. For weak and intermediate interactions, momentum peaks are determined primarily by the underlying lattice vectors, reflecting lattice-dominated quasicrystallinity. At stronger coupling (e.g., $g_{12}=16$), a second ring of eight dominant peaks emerges at smaller momenta, while the original outer ring remains visible. These inner peaks correspond to longer-wavelength real-space modulations induced by local phase separation. The coexistence of the two momentum-space rings thus signals a crossover from a lattice-controlled to an interaction-dominated quasicrystalline regime.

Next, we consider a density-imbalanced mixture at fixed total average density, where component~1 contributes a fraction $0.1$ of the total density and component~2 the remaining fraction $0.9$. Focusing on the repulsive regime with $g_{12} > g$, panels~(a) and (b) of Fig.~\ref{fig:fig2}~ show that for $g_{12} = 15.15$ and $15.3$ (with $g = 15$), the two components coexist and form partially miscible density clusters arranged in an aperiodic pattern exhibiting clear eightfold rotational symmetry.

The density profile along $y=0$, corresponding to Fig.~\ref{fig:fig2}~(a), clearly reveals aperiodic density modulations similar to those observed in experiments employing eightfold quasiperiodic optical lattices~\cite{localization2020}. Increasing the intercomponent repulsion to $g_{12}=15.9$ drives a transition from the partially miscible eightfold quasicrystalline state to a fully phase-separated configuration, as shown in panel~(c). In this regime, the minority component is expelled from the central region and accumulates near the edges of the circular trap, while the majority component occupies the inner region and develops a periodic modulation along $y=0$. This spatial segregation marks the breakdown of the partially miscible quasicrystalline order, with the eightfold rotational symmetry abruptly disappearing in momentum space, as evident in Fig.~\ref{fig:fig2}~(f).
	
Comparing balanced and imbalanced mixtures highlights a fundamental difference in how the two systems respond to increasing intercomponent interactions. In both cases, sufficiently strong coupling initially drives the system into a globally phase-separated state, accompanied by the disappearance of quasicrystalline order, as shown for the balanced mixture in Fig.~\ref{fig:fig3}(g) and for the imbalanced mixture in Fig.~\ref{fig:fig2}(c).

Upon further increasing the intercomponent interaction, the two components exhibit qualitatively different behavior. In the balanced mixture, an unexpected re-emergence of quasicrystalline order occurs: the density patterns reorganize into a locally phase-separated \emph{metastable} configuration (Fig.~\ref{fig:meta_stable}(f)), and the corresponding momentum distribution (Fig.~\ref{fig:meta_stable}(g)) recovers a pronounced eightfold rotational symmetry. This restoration of symmetry is driven by local phase separation, in which alternating domains of comparable density generate longer-wavelength modulations that stabilize quasicrystalline order in a long-lived metastable state rather than in the true ground state.

In sharp contrast, the imbalanced mixture shows no such recovery. Even at larger values of $g_{12}$, the system remains globally phase-separated, and the quasicrystalline peaks do not reappear in momentum space. The minority component is too dilute to form the alternating local domains required to reconstruct eightfold ordering. 
These results demonstrate that population balance plays a crucial stabilizing role. Only in balanced binary condensates can strong intercomponent repulsion induce a re-entrant quasicrystalline phase. In contrast, in imbalanced systems, quasicrystalline order disappears irreversibly once global phase separation occurs.

To quantify the spatial overlap between the two condensates in the balanced and imbalanced case, we define the normalized overlap parameter
\begin{equation}
	\label{eq:overlap}
	\mathcal{O} = 
	\frac{\displaystyle \int d\mathbf{r}\, |\Psi_{1}(\mathbf{r})|^{2}\,|\Psi_{2}(\mathbf{r})|^{2}}
	{\sqrt{\left(\int d\mathbf{r}\, |\Psi_{1}(\mathbf{r})|^{4}\right)
			\left(\int d\mathbf{r}\, |\Psi_{2}(\mathbf{r})|^{4}\right)}}.
\end{equation}
By definition, $\mathcal{O}=0$ corresponds to completely separated components, while larger values indicate stronger spatial mixing.  

Figures~\ref{fig:phase-balance} and \ref{fig:phase-imbalance} show the phase diagrams obtained by evaluating $\mathcal{O}$ as a function of the lattice depth $U$ and the intercomponent interaction strength $g_{12}$.

In the balanced mixture (Fig.~\ref{fig:phase-balance}), the black square marks the parameters used in Fig.~\ref{fig:fig3}(e), corresponding to the transition from QC I to the globally phase-separated state. The black star indicates the parameters of Fig.~\ref{fig:fig3}(f), where the overlap vanishes ($\mathcal{O}=0$), signaling full phase separation. The black circle denotes the parameters of Fig.~\ref{fig:meta_stable}(f), corresponding to QC II, in which density modulations reappear and the momentum distribution restores eightfold rotational symmetry despite local phase separation.

In the imbalanced mixture (Fig.~\ref{fig:phase-imbalance}), two distinct regimes are observed: a quasicrystalline (QC) phase, in which the components remain partially miscible and occupy overlapping regions, and an immiscible phase, where the minority component is expelled toward the edges of the trap. The black square indicates the parameters of Fig.~\ref{fig:fig2}(b), where the density self-organizes into an aperiodic pattern exhibiting eightfold rotational symmetry. The black circle corresponds to Fig.~\ref{fig:fig2}(c), representing a fully phase-separated state in which the eightfold symmetry is lost and the momentum distribution no longer displays quasicrystalline features.
	
\section{Real-time Evolution of Dynamics}
\begin{figure*}[tb]
	\includegraphics[scale=1.5]{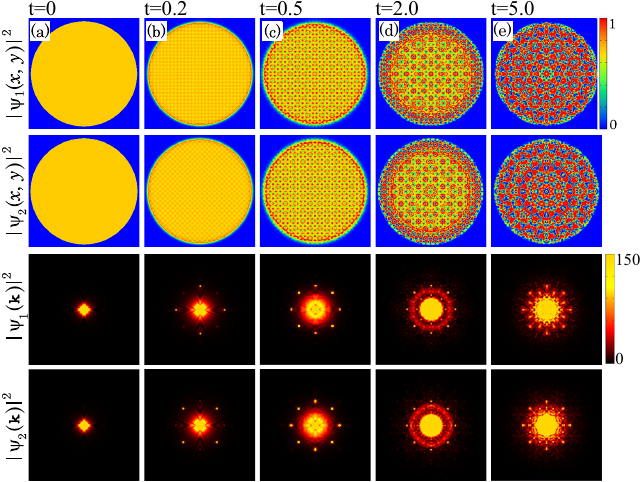}
	\caption{
		Real-time evolution of a balanced two-component BEC in a spin-dependent periodic lattice. The system is initialized in a uniform state $|\Psi_j| = 1/\sqrt{2}$ with small random fluctuations, at an intercomponent interaction strength of $g_{12}=16.5$. All other parameters and the sizes of the real- and momentum-space panels are identical to those in Fig.~\ref{fig:fig3}.
	}
	\label{fig:dyna_bal}
\end{figure*}

\begin{figure*}[tb]
	\includegraphics[scale=1.5]{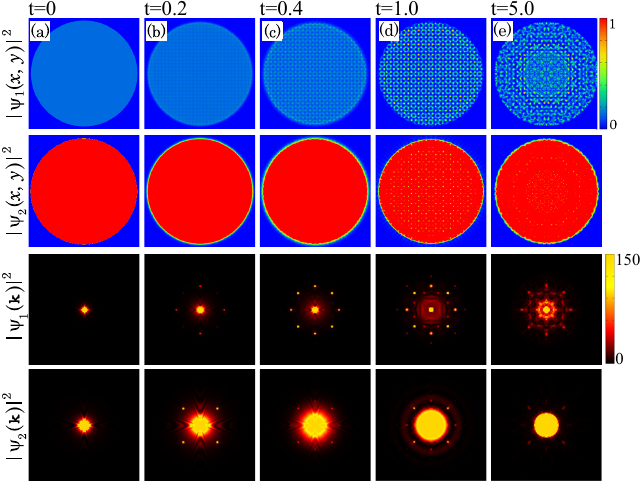}
	\caption{
		Real-time evolution of an imbalanced two-component BEC in a spin-dependent periodic lattice, initialized from a flat state $|\Psi_j| = 1/\sqrt{2}$ with small random fluctuations. Component~1 is the minority and component~2 the majority, with $g_{12}=15.15$. All other parameters and the real- and momentum-space domains are identical to those in Fig.~\ref{fig:fig3}.
	}
	\label{fig:fig4}
\end{figure*}

We analyze the real-time dynamics of binary BECs in spin-dependent periodic lattices, initialized in uniform states $\Psi_j = 1/\sqrt{2}$ with small random perturbations. In balanced mixtures with strong intercomponent interactions ($g_{12}=16.5$, Fig.~\ref{fig:dyna_bal}), initial fluctuations develop into four lattice-induced peaks in momentum space by $t=0.2$, with low-intensity secondary peaks emerging diagonally in component~1 and off-diagonally in component~2. By $t=0.5$, these secondary peaks grow in intensity and, together with the primary lattice-induced peaks, form a clear eightfold rotationally symmetric pattern, which persists at later times despite local phase separation, although the outer peaks slightly decrease in intensity. In imbalanced mixtures ($g_{12}=15.15$, Fig.~\ref{fig:fig4}), the minority component develops eight prominent peaks by $t=0.4$, while the majority shows four dominant peaks with weaker secondary peaks. By $t=1.0$, both components exhibit eightfold symmetry and aperiodic density clusters, but at $t=5.0$, the symmetry is partially distorted as minority peaks weaken and majority outer peaks reduce to four. This comparison demonstrates that quasicrystalline order is robust in balanced mixtures but more fragile in imbalanced ones, highlighting the stabilizing role of population balance.

\section{Conclusion}

We have investigated the formation of QC ground states in two-component BEC loaded into spin-dependent periodic optical lattices. The interplay between intercomponent interactions and component-dependent lattice potentials produces aperiodically modulated density patterns, with the miscible--immiscible transition strongly shaping the overall structure.

In balanced mixtures, weak intercomponent interactions give rise to fourfold symmetry in momentum space for each component, reflecting the underlying lattice geometry and the relative twist of $\pi/4$. As $g_{12}$ increases, secondary peaks appear diagonally in component~1 and off-diagonally in component~2. These peaks combine with the four lattice-induced peaks to form eightfold rotational symmetry, signaling the emergence of quasicrystalline order. At stronger interactions, global phase separation suppresses this symmetry in momentum space. Upon further increasing $g_{12}$, local phase separation develops, the density patterns re-emerge, and the eightfold symmetry is restored in a long-lived metastable state, demonstrating the robustness of quasicrystalline order in balanced mixtures.

In imbalanced mixtures, partially miscible density clusters form at intermediate $g_{12}$, arranging into eightfold rotationally symmetric aperiodic patterns. At larger intercomponent interactions, global phase separation occurs and the quasicrystalline signatures vanish permanently. Real-time simulations confirm that these aperiodic structures are dynamically stable and experimentally accessible.

In contrast to single-component BECs, where quasicrystalline order typically requires explicitly aperiodic potentials with fivefold, eightfold, or higher rotational symmetry, our results demonstrate that binary condensates in spin-dependent periodic lattices can spontaneously form quasicrystalline phases without any aperiodic confinement. The competition between intercomponent interactions and external periodic potentials drives this emergent order, providing a highly tunable platform to explore quantum quasicrystals and their fundamental properties, including excitation spectra~\cite{Piazza}, quantum transport~\cite{Xing2022}, and potential applications in quantum simulation and computation~\cite{Irwin2022}.

\begin{acknowledgments}
	This work was supported by the National Key Research and Development Program of China (Grant No.~2021YFA1401700), the Quantum Science and Technology-National Science and Technology Major Project (Grant No.~2024ZD0300600), Shaanxi Academy of Fundamental Sciences (Mathematics, Physics) (Grant No.~22JSY036). Y.C.~Z. acknowledges the support of the Xiaomi Young Talents program, Xi'an Jiaotong University through the ``Young Top Talents Support Plan" and Basic Research Funding, as well as the High-performance Computing Platform of Xi'an Jiaotong University for the computing facilities. H. S. acknowledges the support by JSPS KAKENHI Grant No.~JP23K03276.
\end{acknowledgments}

\end{document}